\documentclass[conference]{IEEEtran}
\IEEEoverridecommandlockouts

\usepackage{comment}
\usepackage{booktabs} 
\usepackage{hyperref}
\usepackage{url}
\usepackage{fancyhdr}
\usepackage{comment}
\usepackage{graphicx}
\usepackage{caption}
\usepackage{multirow}
\usepackage[scientific-notation=true, round-precision=2, round-mode=figures]{siunitx}
\usepackage{array}
\usepackage{color}
\usepackage{amsmath}
\usepackage{mathtools}
\usepackage[ruled,vlined]{algorithm2e}
\usepackage[framemethod=tikz]{mdframed}
\usepackage{subfigure}

\usepackage[shortlabels]{enumitem}
\usepackage{booktabs}
\usepackage{wasysym}
\usepackage{multirow}
\usepackage{threeparttable}
\usepackage{siunitx}



\newcommand{\sys}{\mbox{\textsc{Janus}}\xspace}

\def\BibTeX{{\rm B\kern-.05em{\sc i\kern-.025em b}\kern-.08em
    T\kern-.1667em\lower.7ex\hbox{E}\kern-.125emX}}
\begin{document}

\title{\sys: Resilient and Adaptive Data Transmission for Enabling Timely and Efficient Cross-Facility Scientific Workflows
}

\author{\IEEEauthorblockN{1\textsuperscript{st} Vladislav Esaulov}
\IEEEauthorblockA{\textit{Georgia State University} \\
Atlanta, GA \\
vesaulov1@student.gsu.edu}
\and
\IEEEauthorblockN{2\textsuperscript{nd} Jieyang Chen}
\IEEEauthorblockA{\textit{University of Oregon} \\
Eugene, OR \\
jieyang@uoregon.edu}
\and
\IEEEauthorblockN{3\textsuperscript{rd} Norbert Podhorszki}
\IEEEauthorblockA{\textit{Oak Ridge National Laboratory} \\
Oak Ridge, TN \\
pnorbert@ornl.gov}
\and
\IEEEauthorblockN{4\textsuperscript{th} Fred Suter}
\IEEEauthorblockA{\textit{Oak Ridge National Laboratory} \\
Oak Ridge, TN \\
suterf@ornl.gov}
\and
\IEEEauthorblockN{5\textsuperscript{th} Scott Klasky}
\IEEEauthorblockA{\textit{Oak Ridge National Laboratory} \\
Oak Ridge, TN \\
klasky@ornl.gov}
\and
\IEEEauthorblockN{6\textsuperscript{th} Anu G Bourgeois}
\IEEEauthorblockA{\textit{Georgia State University} \\
Atlanta, GA  \\
abourgeois@gsu.edu}
\and
\IEEEauthorblockN{7\textsuperscript{th} Lipeng Wan}
\IEEEauthorblockA{\textit{Georgia State University} \\
Atlanta, GA  \\
lwan@gsu.edu}
}

\maketitle

\begin{abstract}


In modern science, the growing complexity of large-scale scientific projects has led to an increasing reliance on cross-facility scientific workflows, where resources and expertise from multiple institutions and geographic locations are leveraged to accelerate scientific discovery. These workflows often require transmitting huge amounts of scientific data through wide-area networks. Although high-speed networks like ESnet and transfer services such as Globus have improved data mobility, several challenges remain. The sheer volume of data can overwhelm network bandwidth, widely used transport protocols such as TCP suffer from inefficiencies due to retransmissions triggered by packet loss, and existing fault-tolerance mechanisms like erasure coding introduce substantial overhead.

In this paper, we propose \sys, a resilient and adaptable data transmission approach designed for cross-facility scientific workflows. Unlike traditional TCP-based methods, \sys leverages UDP, integrates erasure coding for fault tolerance, and combines it with error-bounded lossy compression to reduce overhead.
This novel design allows users to balance data transmission time and accuracy, optimizing transfer performance based on specific scientific requirements. Additionally, \sys dynamically adjusts erasure coding parameters in response to real-time network conditions, ensuring efficient data transfers even in fluctuating environments. We develop optimization models for determining ideal configurations and implement adaptive data transfer protocols to enhance reliability. Through extensive simulations and real-network experiments, we demonstrate that \sys significantly improves transfer efficiency while maintaining data fidelity.
\end{abstract}

\section{Introduction}
\label{sec:intro}
As scientific discovery increasingly relies on advanced technologies and instruments, vast amounts of data are being generated at an unprecedented rate. For example, the International Thermonuclear Experimental Reactor (ITER)~\cite{Iter}, a project designed to demonstrate fusion energy as a large-scale, carbon-free energy source, is expected to produce 2 petabytes of data daily starting in 2035. Similarly, large-scale simulations such as the Nyx cosmology code~\cite{Nyx}, used to simulate the evolution of matter in the universe, can generate several terabytes of data every hour on the fastest U.S. supercomputers. While much of this data is analyzed locally, it must often be transferred across wide-area networks (WANs) to facilitate global collaborations. This is particularly true for ITER, which enables scientists from around the world to conduct fusion experiments remotely and then sends the resulting data from its location in France to the distant researchers for further analysis. This data must sometimes travel across continents. In cases requiring significant computational power, the data may need to be migrated to HPC systems, such as the U.S. Department of Energy’s supercomputers, to accelerate processing. These cross-facility workflows are not only vital for solving complex scientific problems, but also helps facilitate data democratization—the principle that scientific data should be accessible to a wide array of researchers, regardless of their geographic location, institutional affiliation, or available resources, fostering innovation through diverse perspectives. This underscores the critical importance of resilient, efficient, and timely data transmissions. 

Despite advancements in high-speed computer network infrastructures like the Energy Sciences Network (ESnet)~\cite{ESnet} and data transmission services such as Globus~\cite{Globus}, several challenges remain. First, the immense volume of scientific data makes WAN transfers time-consuming, hindering the efficiency of scientific discovery.
Second, widely used transport protocols like TCP can significantly reduce transfer efficiency, particularly over long-distance, high-latency networks due to retransmissions compensating packet loss. As data sizes and transfer distances grow, retransmissions increase delays, undermining the real-time requirements of scientific workflows. Third, while erasure coding can provide fault tolerance and reduce retransmissions, it adds transmission overhead. Fixed fault-tolerance configurations are also ill-suited to dynamic networks with fluctuating packet loss, often resulting in suboptimal performance and further inefficiencies.

In this paper, we propose \sys, a resilient and adaptive data transmission approach designed to enable timely and efficient cross-facility scientific workflows. To accelerate data transmission over WANs, \sys utilizes the UDP protocol instead of TCP. Since UDP lacks a built-in retransmission mechanism to recover from packet loss, \sys incorporates erasure coding to provide fault tolerance. 
To minimize the overhead of erasure coding, \sys incorporates error-bounded lossy compression, a technique already in use by scientists to reduce data size without exceeding acceptable error bounds~\cite{ainsworth2019multilevel2, liang2018error}. This multi-faceted approach allows \sys to balance data transmission time and accuracy.
Specifically, \sys can address two types of user requirements by solving different optimization problems: 1) If the user requires data with a guaranteed error bound, \sys minimizes data transmission time. 2) If the user needs data within a specified time frame, \sys minimizes the error in the received data. Additionally, as network conditions such as packet loss rate can fluctuate during transmission, applying a static level of redundancy to the data cannot achieve optimal transfer efficiency. Therefore, \sys employs adaptive algorithms that dynamically adjust erasure coding parameters based on real-time network conditions, ensuring more efficient data transmission. 

The contributions of this paper are as follows: 1) We develop two optimization models that determine the optimal erasure coding configurations to meet user-defined requirements for both data transmission time and accuracy. These models provide a systematic approach to balancing reliability and efficiency. 2) We design and implement two adaptive data transfer protocols capable of adjusting erasure coding parameters dynamically based on real-time variations in packet loss rates. These protocols ensure optimal performance by minimizing both transmission time and data loss. 3) We present extensive evaluations through simulations and real-world experiments that demonstrate the effectiveness of our approach in improving the reliability and efficiency of scientific data transfers.
\section{Background}
\label{sec:background}

\subsection{Fault Tolerant Data Transmission}


Fault-tolerant data transmission across wide-area networks ensures reliable communication despite challenges like packet loss, corruption, or congestion. TCP, a widely used transport protocol, employs retransmission mechanisms for reliability. It tracks packets with sequence numbers and acknowledgments; if an acknowledgment is not received within a timeout, TCP retransmits the missing packet. However, retransmissions can cause significant delays and variability~\cite{kumar1998comparative, lakshman2000tcp, barakat2000tcp}, making TCP unsuitable for low-latency or time-sensitive scenarios.

Checksums~\cite{stone1998performance, stone2000crc, plummer1989tcp} are another key component of fault-tolerant systems, used to detect errors in transmitted data. A checksum is computed and sent with the data packet; the receiver recalculates it to check for mismatches. If an error is found, retransmission is initiated if supported. While effective for error detection, checksums do not correct errors and rely on additional mechanisms like retransmissions for recovery.

Erasure coding~\cite{rizzo1997effective, srouji2011rdts, leong2012erasure} offers an alternative approach to fault tolerance in data transmission by proactively correcting errors. It divides data into fragments and adds parity fragments. Even if some fragments are lost, the original data can be reconstructed from the remaining fragments, reducing reliance on retransmissions and improving performance in high-loss networks. However, erasure coding introduces computational overhead for encoding and decoding and increases data volume due to the added parity fragments.

\subsection{Scientific Data Refactoring}

Unlike traditional lossy compressors~\cite{ballester2019tthresh, chen2014numarck, lakshminarasimhan2013isabela, lindstrom2006fast, sasaki2015exploration, lindstrom2014fixed, ainsworth2018multilevel, ainsworth2019multilevel, ainsworth2019multilevel2, liang2018error} that typically adhere to a single user-defined error bound, data refactoring decomposes data into multiple components, allowing flexible reconstruction at different approximation levels. For instance, JPEG2000~\cite{rabbani2002book} uses block-based wavelet transformations to support progressive data reconstruction. Similarly, ZFP~\cite{lindstrom2014fixed} provides progressive reconstruction but is limited to a fixed-rate mode. Progressive reconstruction has also been studied for cost-accuracy trade-offs in scientific visualizations~\cite{kumar2014efficient, pascucci2001global, laney2004progressive, paoluzzi2004progressive, shamir2001temporal}, though many approaches lack guarantees on error bounds that are critical for preserving scientific features.

pMGARD~\cite{liang2021error} extends MGARD’s multilevel decomposition approach~\cite{ainsworth2018multilevel, ainsworth2019multilevel, ainsworth2019multilevel2} to offer an error-controlled data refactoring framework for scientific data. It leverages multigrid-based decomposition and $L^2$ projection to break data into multiple levels, each represented by multilevel coefficients used for reconstruction. The reconstruction error is mathematically bounded, ensuring precise error control for each coefficient. pMGARD employs bitplane encoding for multilevel coefficients, optimizing them into a hierarchical structure for progressive reconstruction. This hierarchical approach allows for efficient reconstruction with error bounds determined by the number of levels used, making pMGARD highly suitable for scientific applications.
\section{Optimization in Scientific Data Transmission}
\label{sec:model}

This section provides an overview of \sys and introduces two optimization models that enable it to determine the optimal redundancy for data transmission. All mathematical symbols used in the models are listed in~\autoref{tab:math-symbols}. 

\begin{table}[ht]
  \centering
  \scalebox{0.81}{
    \begin{tabular}{|c|l|}
      \hline
      \textbf{Symbol} & \multicolumn{1}{c|} {\textbf{Description}} \\
      \hline
      $t$ & Latency of transmitting one fragment from sender to receiver\\ 
      \hline
      $r$ & Number of fragments that can be transmitted per second\\ 
      \hline
      $\lambda$ & Packet loss rate (number of packet losses per second)\\ 
      \hline
      $n$ & Number of fragments in each fault-tolerant group\\ 
      \hline
      $L$ & Total number of levels after refactoring\\
      \hline
      $l$ & Total number of levels sent by the sender\\
      \hline
      $k_i$ & Number of data fragments in each fault-tolerant group of level $i$\\ 
      \hline
      $m_i$ & Number of parity fragments in each fault-tolerant group of level $i$\\ 
      \hline
      $s$ & Size of each fragment\\ 
      \hline
      $S_i$ & Size of each level after data refactoring\\ 
      \hline
      \multirow{2}{*}{$\varepsilon_{i}$} & Relative L-infinity error when using level $1, 2, \ldots, i$  \\
      & of the refactored data to reconstruct the original data\\
      \hline  
      $N_i$ & Number of fault-tolerant groups of level $i$\\ 
      \hline
    \end{tabular}
  }
\caption{Mathematical symbols used in our models}
\vspace{-0.5cm}
  \label{tab:math-symbols}
\end{table}

\subsection{An Overview of \sys}

To enable efficient transmission of large scientific datasets over wide-area networks, \sys utilizes the low latency of the UDP protocol for rapid data transfer and addresses packet loss with erasure coding and error-bounded lossy compression, allowing recovery without costly retransmissions.

\begin{figure}[h]
  \centering
  \subfigure[Sender: The original data is refactored into a hierarchical representation, with each level further divided into fixed-size data fragments. Erasure coding is applied to every $k$ data fragments to generate $m$ parity fragments, forming fault-tolerant groups of $k+m$ fragments. All fragments are transmitted via UDP, with each fragment encapsulated in a separate UDP packet.]{
    \label{fig:send}
    \includegraphics[width=0.97\columnwidth]{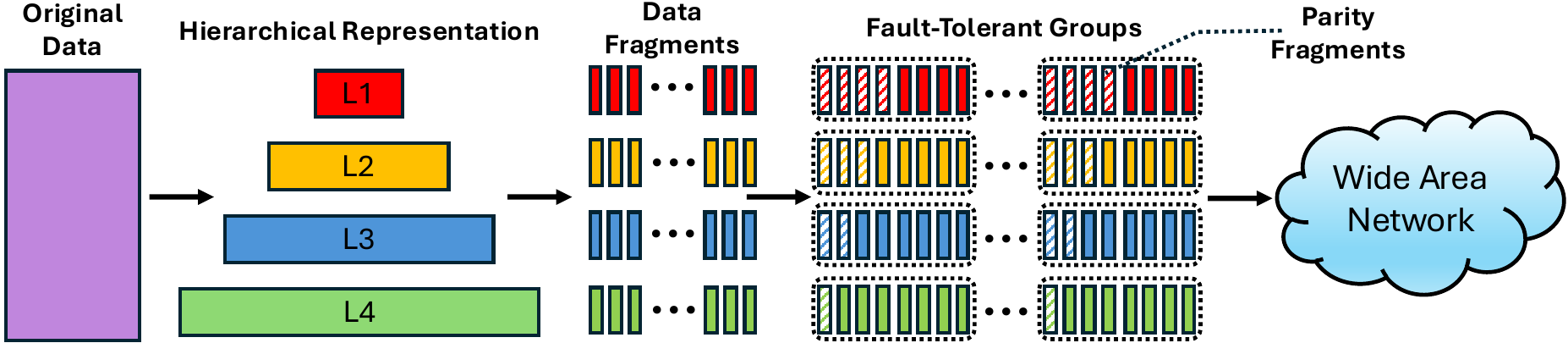}}
  \subfigure[Receiver: UDP packets may be lost during transmission. In each fault-tolerant group, as long as no more than $m$ fragments are lost, the missing data fragments can still be recovered. However, if more than $m$ fragments are lost, at least one data fragment becomes unrecoverable, leading to corruption of the entire level. In such cases, an approximation of the original data is reconstructed using the remaining hierarchical structure.]{
    \label{fig:receive}
    \includegraphics[width=0.97\columnwidth]{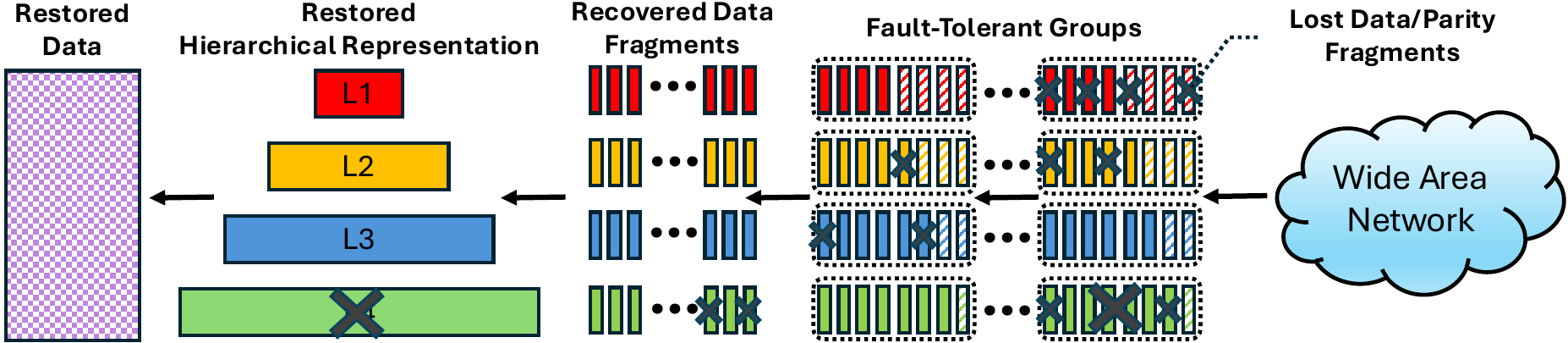}}
  \caption{An illustration of data transfer with \sys}
  \label{fig:example}
\end{figure}

To illustrate how \sys works, consider the example in~\autoref{fig:example}. As shown in~\autoref{fig:send}, on the sender side, \sys uses pMGARD to refactor the original data into a hierarchical structure with four levels. The sizes of these four levels are denoted by $S_1$, $S_2$, $S_3$, and $S_4$, where $S_1 < S_2 < S_3 < S_4$. 
Adding an additional level from top to bottom improves the reconstruction accuracy, with the error bounds decreasing as follows: $\varepsilon_{1}$ for level 1, $\varepsilon_{2}$ for levels 1 and 2, $\varepsilon_{3}$ for levels 1–3, and so on, where $\varepsilon_{1} > \varepsilon_{2} > \ldots > \varepsilon_{L}$.
The relative L-infinity error, defined as
\begingroup
\setlength{\abovedisplayskip}{3pt}
\setlength{\belowdisplayskip}{3pt}
\begin{equation}
\label{rel-l-inf-err}
\varepsilon = \frac{\max(\| d - \tilde{d} \|)}{\max(\| d \|)}, 
\end{equation}
\endgroup
measures the error between the original data $d$ and the reconstructed data $\tilde{d}$, where $\max(| d - \tilde{d} |)$ is the maximum absolute difference and $\max(| d |)$ is the maximum value in the original data. In the worst-case scenario, when no levels are available for reconstruction, the relative error is 1.
Each level is divided into fragments, with erasure coding applied: for every $k$ data fragments, $m$ parity fragments are generated. The total number of fragments, $n = k + m$, remains constant across levels. For example, $k_1=4$, $m_1=4$; $k_2=5$, $m_2=3$; $k_3=6$, $m_3=2$; and $k_4=7$, $m_4=1$. These data and parity fragments form fault-tolerant groups (FTGs) for each level. All fragments are transmitted via UDP, with each fragment encapsulated in its own UDP packet.

During transmission, some packets may be lost, but parity fragments allow the receiver to recover many lost fragments without retransmission. As shown in~\autoref{fig:receive}, at level 1, data can be fully recovered as long as no more than 4 fragments are lost per FTG. Similarly, lost fragments at levels 2 and 3 can be recovered using parity fragments. At level 4, however, losing 2 fragments from an FTG cannot be recovered due to the single parity fragment. Without retransmission, only the top three levels can be reconstructed within the error bound $\varepsilon_{3}$. The number of parity fragments per level can be adjusted according to user requirements.

\subsection{Optimization Models}
\label{sec:optimization_models}

\sys is designed to address two types of user requirements by solving the following two optimization models. 

\subsubsection{Minimizing Data Transmission Time with a Guaranteed Error Bound on the Received Data} 
Scientists often require that received data maintain an error within a specified bound after decompression to ensure compression-induced errors do not compromise downstream analyses. If the user-specified bound is $\varepsilon$ and $\varepsilon_l \leq \varepsilon < \varepsilon_{l-1}$, \sys ensures that at least the first $l$ levels of the refactored data are transferred. Lost fragments can be recovered as long as their number does not exceed the parity fragments within each FTG. If insufficient levels are recoverable, \sys performs passive retransmission of FTGs with unrecoverable losses, which may require multiple rounds due to potential additional losses during retransmission.

To minimize overall transmission time, we develop an optimization model. The total size of the first $l$ levels is $S = S_1 + \dots + S_l$, and the number of FTGs is $N = \frac{S}{(n-m)s}$. The initial transmission time is $t + \frac{nN-1}{r}$, and expected retransmission time depends on the number of FTGs requiring retransmission per round.
For low packet loss rates, unrecoverable losses can be modeled as independent events, where the number of FTGs with unrecoverable losses follows a binomial distribution.
If $N$ FTGs are transmitted initially and the probability of unrecoverable loss in a single FTG is $p$, the expected number of FTGs requiring retransmission in the first round is $\sum_{K=0}^{N} K \binom{N}{K} p^K (1-p)^{N-K} = Np$. Consequently, the expected number of FTGs needing retransmission in the $i$-th round is $Np^i$. 
The probability that at least one FTG in the $i$-th round experiences unrecoverable loss is $1-(1-p)^{Np^i}$. Thus, the expected total transmission time is
\begingroup
\setlength{\abovedisplayskip}{3pt}
\setlength{\belowdisplayskip}{3pt}
\begin{equation}
\label{ETtotal}
    E[T_{total}] = t+\frac{nN-1}{r}+\sum\limits_{i=1}^{\infty}[1-(1-p)^{Np^{i-1}}](t+\frac{nNp^i-1}{r}) 
\end{equation}
\endgroup
which converges as $i \to \infty$ since $0<p<1$.

Estimating $p$, the probability that an FTG cannot be recovered, depends on the number of fragments $n$, transmission rate $r$, and network latency $t$. If $T = t + (n-1)/r$ is the total time for sending an FTG, then $u = rt + n - 1$ fragments are sent during $T$. Assuming a roughly constant packet loss rate $\lambda$, fragment losses can be modeled using a Poisson distribution.
Specifically, the probability of exactly $j$ out of the $u$ fragments being lost within $T$ is given by:
\begingroup
\setlength{\abovedisplayskip}{3pt}
\setlength{\belowdisplayskip}{3pt}
\begin{equation}
Pr(v = j) = \frac{(\lambda T)^j e^{-\lambda T}}{j!} =\frac{{\lambda (t+\frac{n-1}{r})}^j e^{-\lambda (t+\frac{n-1}{r})}}{j!} 
\end{equation}
\endgroup
For an FTG with $m$ parity fragments, the probability of unrecoverable loss given $j>m$ lost fragments is
\begingroup
\setlength{\abovedisplayskip}{3pt}
\setlength{\belowdisplayskip}{3pt}
\begin{equation}
Pr(\text{unrecoverable}|v = j) = \frac{\sum\limits_{w=m+1}^{n}\binom{n}{w}\binom{u-n}{j-w}}{\binom{u}{j}}
\end{equation}
\endgroup
and the overall probability is
\begingroup
\setlength{\abovedisplayskip}{3pt}
\setlength{\belowdisplayskip}{3pt}
\begin{equation}
\begin{split}
\label{p}
p=&\sum\limits_{j=m+1}^{u}Pr(\text{unrecoverable}|v = j)Pr(v = j) \\
=&\sum\limits_{j=m+1}^{u}\frac{\sum\limits_{w=m+1}^{n}\binom{n}{w}\binom{u-n}{j-w}}{\binom{u}{j}}\frac{{\lambda (t+\frac{n-1}{r})}^j e^{-\lambda (t+\frac{n-1}{r})}}{j!} \\
=&\sum\limits_{j=m+1}^{rt+n-1}\frac{\sum\limits_{w=m+1}^{n}\binom{n}{w}\binom{rt-1}{j-w}}{\binom{rt+n-1}{j}}\frac{{\lambda (t+\frac{n-1}{r})}^j e^{-\lambda (t+\frac{n-1}{r})}}{j!} 
\end{split}
\end{equation}
\endgroup

At high packet loss rates, unrecoverable losses across FTGs are correlated, invalidating the above calculation. We therefore we adopt an alternative approach to calculate $p$. 
Given that 
the average fragment losses each of these FTGs experiences is $\lambda T/(\frac{t+(n-1)/r}{n/r})=\frac{\lambda n}{r}$, if $\frac{\lambda n}{r}>1$ (more than one fragment loss per FTG), 
the probability $p$ that an FTG experiences unrecoverable data losses can be approximately estimated as:
\begingroup
\setlength{\abovedisplayskip}{3pt}
\setlength{\belowdisplayskip}{3pt}
\begin{equation}
\begin{split}
\label{p_big_lambda}
p =&Pr(\text{more than} \ m \ \text{fragment losses in an FTG}) \\
=&1-\sum\limits_{j=0}^{m}\frac{{(\frac{\lambda n}{r})}^j e^{-(\frac{\lambda n}{r})}}{j!} 
\end{split}
\end{equation} 
\endgroup

In summary, the following optimization problem needs to be solved:
\begingroup
\setlength{\abovedisplayskip}{3pt}
\setlength{\belowdisplayskip}{3pt}
\begin{equation}
\begin{split}
\label{ETotal_opt}
\operatorname*{arg\,min}_{ m \in \{0,\ldots,n/2\} } &\quad E[T_{total}] \\
\text{s.t.}& \quad \text{Calculate $p$ using~\autoref{p_big_lambda}, if $\frac{\lambda n}{r}>1$}   \\
& \quad \text{Calculate $p$ using~\autoref{p}, otherwise}
\end{split}
\end{equation}
\endgroup
Regardless of how $p$ is calculated, increasing $m$ has two opposing effects on $E[T_{total}]$: first, since $N = \frac{S}{(n-m)s}$, a larger $m$ increases the number of FTGs, resulting in more fragments transmitted during both initial transfer and retransmissions, which raises $E[T_{total}]$; second, a larger $m$ reduces $p$, lowering the likelihood of retransmissions and offsetting the increase caused by a higher $N$. Consequently, an optimal $m$ exists that minimizes $E[T_{total}]$. In practice, with relatively small $p$, $E[T_{total}]$ converges for $i > 50$. Since $m$ is an integer and must satisfy $m \le n/2$ (the number of parity fragments cannot exceed the number of data fragments), determining the optimal $m$ is straightforward using either an exhaustive search or a lightweight numerical optimization.


\subsubsection{Minimizing Errors in the Received Data with a Guaranteed Data Transmission Time}

In some scenarios, scientists prioritize receiving data quickly within a deterministic timeframe, even if it contains some error, rather than waiting for highly accurate data over a prolonged, unpredictable period. For example, in scientific visualization, rapidly rendering preliminary representations of large datasets allows researchers to identify trends, detect anomalies, or formulate hypotheses, even with imperfect data. To ensure predictable transmission times, retransmissions are minimized, relying primarily on parity fragments from erasure coding. The key challenge is determining which levels to transmit and how many parity fragments to generate for each FTG so total transmission time stays within the user-specified limit while minimizing errors. The following optimization model addresses this challenge.

Let the size of the $j$-th level of refactored data be $S_j$, with $m_j$ parity fragments generated. The number of FTGs for level $j$ is $N_j = \frac{S_j}{(n-m_j)s}$. If the first $l$ of $L$ levels are sent without retransmission, the total transmission time is
\begingroup
\setlength{\abovedisplayskip}{3pt}
\setlength{\belowdisplayskip}{3pt}
\begin{equation}
\begin{split}
T_{total} =& t+\frac{n(N_1+N_2+\ldots+N_l)-1}{r} \\
=& t+\frac{n\sum\limits_{j=1}^{l}\frac{S_j}{(n-m_j)s}-1}{r}
\end{split}
\end{equation}
\endgroup
Given a user-specified time constraint $\tau$ and $0 \le m_j \le n/2$, the feasible range of $l$ can be estimated by
\begingroup
\setlength{\abovedisplayskip}{3pt}
\setlength{\belowdisplayskip}{3pt}
\begin{equation}
\label{l_range}
  t+\frac{n\sum\limits_{j=1}^{l}\frac{S_j}{(n-0)s}-1}{r} \leq T_{total} \leq t+\frac{n\sum\limits_{j=1}^{l}\frac{S_j}{(n-\frac{n}{2})s}-1}{r} \leq \tau
\end{equation}
\endgroup

To estimate errors in the received data, let $p_1$ denote the probability that an FTG in the first level experiences unrecoverable data loss. Assuming such losses are independent across groups, the probability that the first level cannot be recovered is $1-(1-p_1)^{N_1}$. Since the first level is essential for data reconstruction, its loss results in a relative error of $\varepsilon_0 = 1$. If the first level is successfully recovered but the second level is not, the data can still be reconstructed with an error of $\varepsilon_1$, occurring with probability $(1-p_1)^{N_1}[1-(1-p_2)^{N_2}]$. Similarly, if the first two levels are recovered but the third fails, the error is $\varepsilon_2$, with probability $(1-p_1)^{N_1}(1-p_2)^{N_2}[1-(1-p_3)^{N_3}]$. Following this pattern, for  $l$ transmitted levels, the expected error of the reconstructed data is:
\begingroup
\setlength{\abovedisplayskip}{3pt}
\setlength{\belowdisplayskip}{3pt}
\begin{equation}
\begin{split}
\label{epsilon}
    E[\varepsilon]=&[1-(1-p_1)^{\frac{S_1}{(n-m_1)s}}]\varepsilon_0+ \\
                   &\sum\limits_{i=2}^{l-1}\prod\limits_{j=1}^{i-1}(1-p_j)^{\frac{S_j}{(n-m_j)s}}[1-(1-p_i)^{\frac{S_i}{(n-m_i)s}}]\varepsilon_{i-1}+ \\
                   &\prod\limits_{j=1}^{l}(1-p_j)^{\frac{S_j}{(n-m_j)s}}\varepsilon_{l}
\end{split}
\end{equation}
\endgroup
As previously discussed, the method for calculating $p_j$ depends on the packet loss rate. When the packet loss rate is relatively low, $p_j$ is calculated using~\autoref{p}. Conversely, when the packet loss rate is high and correlations between FTGs become significant, $p_j$ is determined using~\autoref{p_big_lambda}. 

In summary, the complete optimization model, combining all components, is formulated as a nonlinear integer programming problem as follows:
\begingroup
\setlength{\abovedisplayskip}{3pt}
\setlength{\belowdisplayskip}{3pt}
\begin{equation}
\begin{split}
\label{epsilon_opt}
\operatorname*{arg\,min}_{m_{j}, \forall j \in \{1,\ldots,l\}} &  [1-(1-p_1)^{\frac{S_1}{(n-m_1)s}}]\varepsilon_0+ \\
                   &\sum\limits_{i=2}^{l-1}\prod\limits_{j=1}^{i-1}(1-p_j)^{\frac{S_j}{(n-m_j)s}}[1-(1-p_i)^{\frac{S_i}{(n-m_i)s}}]\varepsilon_{i-1} \\
                   & +\prod\limits_{j=1}^{l}(1-p_j)^{\frac{S_j}{(n-m_j)s}}\varepsilon_{l} \\
\text{s.t.}& \quad  t+\frac{n\sum\limits_{j=1}^{l}\frac{S_j}{(n-m_j)s}-1}{r} \leq \tau 
\end{split}
\end{equation}
\endgroup
This model can be solved using numerical solvers such as SCIP~\cite{BolusaniEtal2024OO} or other optimization tools capable of handling nonlinear integer programming.

\section{Adaptive Data Transfer Protocols}
\label{sec:algorithm}

In real-world networks, packet loss rates often fluctuate over time, 
making fault-tolerance configurations based on constant loss rates suboptimal and potentially degrading transmission efficiency or violating the user-specified constraints on total transmission time. To address this, we incorporate adaptive mechanisms into the data transfer protocols, dynamically adjusting fault-tolerance settings in response to changing loss rates to consistently satisfy the two user requirements outlined in~\autoref{sec:optimization_models}.

\subsection{Data Transfer with Guaranteed Error Bound}

The protocol for transferring data between sender and receiver with guaranteed error bound is shown in~\autoref{alg:transfer_with_err_bound}.

\begin{algorithm}[h]
\SetAlCapFnt{\small}
\SetAlCapNameFnt{\small}
\caption{Data Transfer with Guaranteed Error Bound}
\label{alg:transfer_with_err_bound}
\scalebox{0.8}{%
  \parbox{\linewidth}{%
\SetKwInput{Input}{Input}
\SetKwBlock{Sender}{Sender}{}
\SetKwBlock{ThreadOne}{Parity Generation Thread}{}
\SetKwBlock{ThreadTwo}{Transmission Thread}{}
\SetKwBlock{Receiver}{Receiver}{}
\SetKw{Break}{break}

\Sender{
\Input{$\varepsilon$, $s$, $t$, $n$, $r_{ec}$, $r_{link}$, $\lambda$}
\ThreadOne{
Determine $l$ such that $\varepsilon_{l}\leq \varepsilon< \varepsilon_{l-1}$\;
Divide $l$ levels into $d=\frac{\sum_{i=1}^{l}S_i}{s}$ data fragments\;
Compute $r=min\{r_{ec}, r_{link}\}$\;
Solve~\autoref{ETotal_opt} to obtain $m$\;
\ForEach{$k$ ($k=n-m$) data fragments}{
Generate $m$ parity fragments to form an FTG\;
Store data and parity fragments of this FTG to $buf$\;
\If{updated $\lambda$ notified by Receiver}{
Update $m$ by solving~\autoref{ETotal_opt} again\;
}
}
}
\ThreadTwo{
\While{$d>0$}{
Transmit fragments of an FTG in $buf$\;
$d \gets d-k$\;
}
Notify Receiver transmission ended\;
\While{True}{
Wait for $lost\_FTGs\_list$ from Receiver\;
\If{$lost\_FTGs\_list$ is empty}{
\Break
}
\ForEach{$lost\_FTG$ in $lost\_FTGs\_list$}{
Obtain fragments of $lost\_FTG$ from $buf$ and transmit\;
}
Notify Receiver transmission ended\;
}

}
}
\Receiver{
\Input{$T_W$}
Initialize $t_{start} \gets t_0$, $lost \gets 0$\;
\While{True}{
Receive fragments from Sender\;
\ForEach{$FTG$} {
\If{fragment loss occurs}{
Increment $lost$ by number of lost fragments\;
\If{data fragment loss occurs}{
\If{number of lost fragments $\leq m$}{
Recover lost data fragments\;
}
\Else{
Add $FTG$ to $lost\_FTGs\_list$\;
}
}
}
}
\If{$t_{now}-t_{start}\geq T_W$}{
Update $\lambda \gets \frac{lost}{T_W}$ and notify Sender\;
$t_{start} \gets t_{now}$, $lost \gets 0$
}
\If{transmission ended notified by Sender}{
Send $lost\_FTGs\_list$ to Sender\;
\If{$lost\_FTGs\_list$ is empty}{
\Break
}
}
}
}
}}
\end{algorithm}

The sender is multithreaded to handle parity generation and data transmission concurrently. The parity generation thread first analyzes the user-specified error bound to determine the levels to transmit, partitions the data into fragments, and calculates the required number of parity fragments ($m$) for each FTG by solving the optimization problem in~\autoref{ETotal_opt}, which depends on the actual packet transmission rate ($r$) and packet loss rate ($\lambda$). The transmission rate $r$ is the minimum of the parity generation rate ($r_{ec}$) and the network link rate ($r_{link}$) for UDP packets of size $s$. The thread continuously generates $m$ parity fragments for every $k$ data fragments while monitoring updates to $\lambda$ from the receiver, dynamically recalculating $m$ as network conditions change. Simultaneously, the transmission thread sends fragments in FTGs, notifies the receiver upon completion, and retransmits lost FTGs iteratively until none remain.

The receiver operates continuously to handle incoming fragments from the sender. For each FTG, it verifies if any fragment loss has occurred and counts the number of lost fragments within a user-specified time window, $T_W$. If data fragment loss is detected and the number of lost fragments in the FTG does not exceed $m$, the receiver recovers the missing data fragments using the remaining data and parity fragments. If recovery is not possible, the FTG is marked as unrecoverable and added to a list of lost FTGs. Once $T_W$ has elapsed, the receiver calculates the updated packet loss rate, $\lambda$ based on the total fragment losses within the time window and notifies the sender with the new $\lambda$. Additionally, when notified by the sender that the transmission has concluded, the receiver provides the sender with the list of lost FTGs for retransmission.

\subsection{Data Transfer with Guaranteed Time}

The protocol for transferring data between sender and receiver with guaranteed transmission time is shown in~\autoref{alg:transfer_with_time}.

\begin{algorithm}[h]
\SetAlCapFnt{\small}
\SetAlCapNameFnt{\small}
\caption{Data Transfer with Guaranteed Time}
\label{alg:transfer_with_time}
\scalebox{0.8}{%
  \parbox{\linewidth}{%
\SetKwInput{Input}{Input}
\SetKwBlock{Sender}{Sender}{}
\SetKwBlock{ThreadOne}{Parity Generation Thread}{}
\SetKwBlock{ThreadTwo}{Transmission Thread}{}
\SetKwBlock{Receiver}{Receiver}{}
\SetKw{Break}{break}

\Sender{
\Input{$\tau$, $[S_1,S_2,\ldots,S_L]$, $[\varepsilon_1,\varepsilon_2,\ldots,\varepsilon_L]$, $s$, $t$, $n$, $r_{ec}$, $r_{link}$, $\lambda$}
\ThreadOne{
Compute $r=min\{r_{ec}, r_{link}\}$\;
Find all possible $l$ values by solving~\autoref{l_range}\;
\If{no possible $l$ is found due to small $\tau$}{
Throw an exception\;
}
$\varepsilon_{min}\gets 1.0$\;
\ForEach{possible $l$ value $l'$}{
Obtain $[m_1,m_2,\ldots, m_{l'}]$ by solving~\autoref{epsilon_opt}\;
$\varepsilon \gets$ value of~\autoref{epsilon}\;
\If{$\varepsilon < \varepsilon_{min}$}{
$\varepsilon_{min} \gets \varepsilon$\; 
$l \gets l'$ and $M \gets [m_1,m_2,\ldots, m_{l'}]$\;
}
}
Calculate the total number of data fragments $d=\frac{\sum_{i=1}^{l}S_i}{s}$\;
\ForEach{$i \in \{1,2,\ldots,l\}$} {
Divide the $i$-th level into $d_i = \frac{S_i}{s}$ data fragments\;
\ForEach{$k_i$ ($k_i=n-m_i$) data fragments}{
Generate $m_i$ parity fragments to form an FTG\;
Store data and parity fragments of this FTG to $buf$\;
\If{updated $\lambda$ notified by Receiver}{
Update $\{m_i,\ldots,m_l\}$ by solving~\autoref{epsilon_opt} again\;
}
}
}
}
\ThreadTwo{
\While{$d>0$}{
Transmit fragments of an FTG in $buf$\;
Extract $m_i$ from the fragments' metadata\;
$d \gets d-(n-m_i)$\;
}
Notify Receiver transmission ended\;
}
}
\Receiver{
\Input{$T_W$}
Initialize $t_{start} \gets t_0$, $lost \gets 0$\;
\While{True}{
Receive fragments from Sender\;
\ForEach{$FTG$} {
Extract $m_i$ from the fragments' metadata\;
\If{fragment loss occurs}{
Increment $lost$ by number of lost fragments\;
\If{data fragment loss occurs}{
\If{number of lost fragments $\leq m_i$}{
Recover lost data fragments\;
}
\Else{
Mark the $i$-th level as unrecoverable\;
}
}
}
}
\If{$t_{now}-t_{start}\geq T_W$}{
Update $\lambda \gets \frac{lost}{T_W}$ and notify Sender\;
$t_{start} \gets t_{now}$, $lost \gets 0$
}
\If{transmission ended notified by Sender}{
\Break
}
}
}
}}
\end{algorithm}

The sender in~\autoref{alg:transfer_with_time} also uses multithreading to handle parity generation and data transmission concurrently. The parity generation thread first calculates the effective packet transmission rate $r$, the minimum of the parity generation rate ($r_{ec}$) and network link rate ($r_{link}$). Using $r$, it solves~\autoref{l_range} to identify feasible numbers of levels that can be transmitted within the deadline $\tau$. If no feasible solution exists, the protocol raises an exception indicating the deadline is too tight. Among feasible options, the thread selects the optimal number of levels $l$ and computes the best redundancy parameters $[m_1, \dots, m_l]$ by solving~\autoref{epsilon_opt}, minimizing overall error $\varepsilon$ while meeting timing constraints. Each level is divided into fragments of size $s$, and for level $i$, $m_i$ parity fragments are generated for every $k_i = n-m_i$ data fragments to form an FTG. The thread dynamically adapts to network changes: when the receiver reports an updated packet loss rate $\lambda$, it recalculates $m_i$ using~\autoref{epsilon_opt}, excluding already transmitted FTGs and adjusting $\tau$ for elapsed time.

The receiver continuously processes incoming fragments and monitors fragment losses. For each FTG, it extracts the level-specific redundancy parameter ($m_i$) from the fragments' metadata to determine recoverability. If data fragments are lost and the total number of lost fragments does not exceed $m_i$, the receiver restores the missing fragments. Otherwise, the affected level is marked as unrecoverable. The receiver also maintains a time window ($T_W$) to calculate the packet loss rate ($\lambda$) and periodically notifies the sender of updated $\lambda$ values. Once the sender signals the end of transmission, the receiver concludes its operations.
\section{Evaluation}
\label{sec:evaluation}

Simulation is the primary evaluation method for the proposed adaptive data transfer protocols, offering a controlled environment to systematically explore protocol behavior under diverse and extreme packet loss conditions, which are difficult to manage in real networks. To complement this, we also performed real network tests to validate the protocols’ functionality in practical scenarios.

\subsection{Testing Data and Network}
The scientific data for our evaluation was generated by the Nyx cosmology simulation~\cite{Nyx} and refactored into a four-level hierarchical representation using the pMGARD~\cite{liang2021error} library. The level sizes were $S_1 = 668$~MB, $S_2 = 2.67$~GB, $S_3 = 5.42$~GB, and $S_4 = 17.99$~GB, with relative L-infinity reconstruction errors $\varepsilon_1 = 0.004$, $\varepsilon_2 = 0.0005$, $\varepsilon_3 = 0.00006$, and $\varepsilon_4 = 0.0000001$. We do not evaluate data refactoring here, as it is typically performed during data generation and stored locally; evaluations of data refactoring techniques can be found in existing works~\cite{liang2021error, wan2023rapids}.
The testing network used two virtual machines on CloudLab~\cite{cloudlab}, part of the National Science Foundation's NSFCloud program. The sender VM was on the d430s cluster at Emulab, University of Utah, while the receiver VM was on the c6420 cluster at Clemson University. Data transfers between the sender and receiver traverse the wide-area network.

\subsection{Simulation-Based Evaluation}

\subsubsection{Design of the Simulation Model} 
We developed a discrete-event simulation framework using SimPy~\cite{simpy}, a Python library for modeling discrete-event systems, to capture key operations of data transfer protocols. The sender continuously generates and transmits packets, producing $m$ parity fragments for every $k$ data fragments and sending each packet every $1/r$ seconds, where $r$ is the transmission rate. Each packet experiences a latency of $t$ seconds, and if not lost, it is added to a FIFO queue. The receiver process retrieves packets from this queue. Packet loss is simulated by generating random intervals between loss events; packets are marked lost if the loss queue is nonempty, after which the queue is cleared. A separate process handles control message exchanges between sender and receiver.

Our framework also extends the core simulation to model specific protocols. For TCP, parity fragment generation is disabled, and acknowledgment and retransmission mechanisms are simulated. For UDP-based protocols with erasure coding, the receiver analyzes metadata to identify FTGs with data losses and checks if recovery is possible. If the required data levels are insufficient and an error bound must be guaranteed, the receiver requests retransmissions until all necessary fragments are received. For the adaptive data transfer protocols, the receiver periodically measures the packet loss rate and updates the sender, which then recalculates the optimal $m$ values and adjusts parity generation to ensure efficient, reliable data transfer.

\subsubsection{Simulation Parameter Settings}

To determine the simulation parameters, we conducted a series of detailed measurements over the testing network to characterize its latency, throughput, and packet loss behavior. We first measured the transmission latency $t$ by continuously sending 4,096-byte UDP data fragments for 15 minutes and recording the latency of each fragment. This experiment was repeated across multiple days to capture potential temporal variations in network performance. The observed latencies were highly stable, ranging from 0.029 to 0.039 seconds, and we therefore used the average value $t = 0.034$ seconds in our simulations.

Next, we evaluated the maximum network transmission rate, $r_{link}$, and the parity fragment generation rate, $r_{ec}$. Using 4,096-byte packets, we measured a link throughput of $r_{link} = 29{,}000$ packets per second. To quantify $r_{ec}$, we used the ``liberasurecode'' library~\cite{liberasurecode} to generate parity fragments for FTGs consisting of $n = 32$ fragments. As the number of parity fragments $m$ increased from 1 to 16, the measured $r_{ec}$ decreased from 319,531 to 41,561 fragments per second due to the growing encoding overhead. Because $r_{ec}$ remained higher than $r_{link}$ even at $m = 16$, the actual transmission rate was bounded by the network link capacity, i.e., $r = r_{link}$. However, in high-bandwidth environments where $r_{link} > r_{ec}$, the effective throughput would instead be limited by the encoding rate $r_{ec}$.
We then characterized packet loss behavior empirically using the same testing network. From these measurements, we identified three representative packet loss intensity levels: low ($\lambda_{low} = 20$ packets per second), medium ($\lambda_{medium} = 400$ packets per second), and high ($\lambda_{high} = 1000$ packets per second). The packet loss process was modeled as a Poisson process, in which the inter-loss intervals follow an exponential distribution parameterized by the rate $\lambda$, capturing the random and memoryless nature of loss events in large-scale networks.

To simulate time-varying packet loss, we further adopted a hidden Markov model (HMM)~\cite{rodbro2006hidden, rossi2006joint, silveira2007predicting} with three latent states—low, medium, and high—corresponding to different congestion conditions. Each state was associated with a Gaussian distribution of packet loss rates: low ($\mu=20$, $\sigma=2$), medium ($\mu=400$, $\sigma=40$), and high ($\mu=1000$, $\sigma=100$). State transitions followed a continuous-time Markov chain with exponentially distributed holding times (rate = 0.04), resulting in an average transition every 25 seconds. Within each state, packet loss events were sampled from the corresponding Gaussian distribution, producing a realistic, temporally correlated pattern of network variability.

Finally, protocol-specific parameters were configured using widely accepted defaults. For TCP, the retransmission timeout (RTO) was set to twice the measured transmission latency, and the duplicate acknowledgment threshold for triggering fast retransmission was set to 3. For the adaptive data transfer protocols, the observation window for estimating and updating the packet loss rate was configured to 3 seconds, providing a balance between responsiveness to rapid network fluctuations and the stability of loss rate estimation.

\subsubsection{Data Transfer with Guaranteed Error Bound}

\begin{figure*}[htbp]
  \centering 
  \subfigure[Low packet loss rate]{ 
  \label{fig:lambda_19_time}
    \includegraphics[width=.23\linewidth]{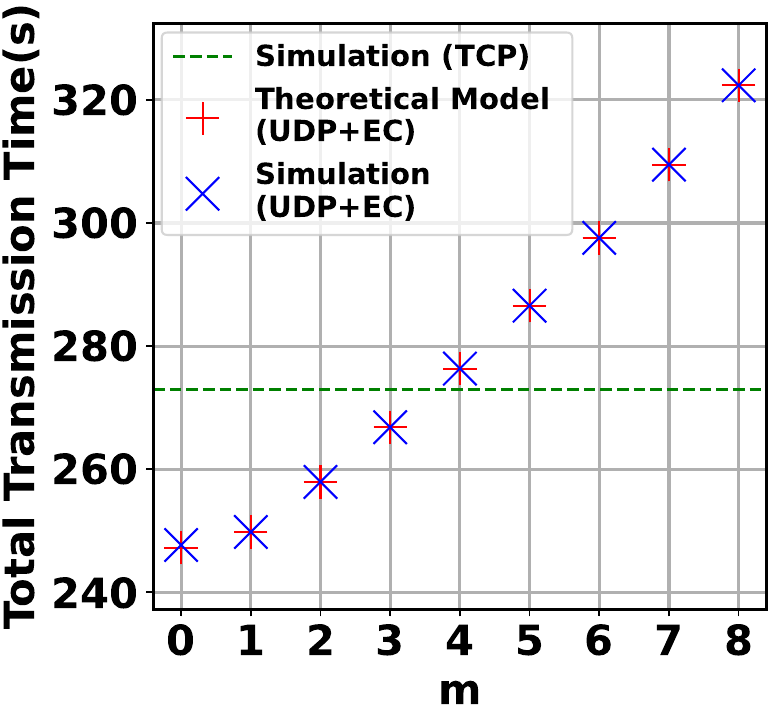}}
  \subfigure[Medium packet loss rate]{ 
  \label{fig:lambda_383_time}
    \includegraphics[width=.23\linewidth]{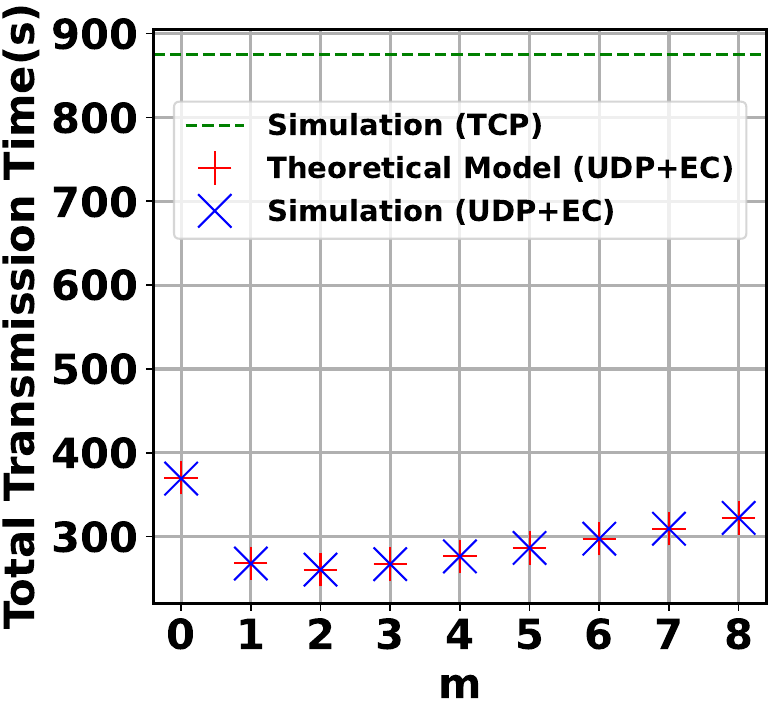}}
  \subfigure[High packet loss rate]{ 
  \label{fig:lambda_957_time}
    \includegraphics[width=.23\linewidth]{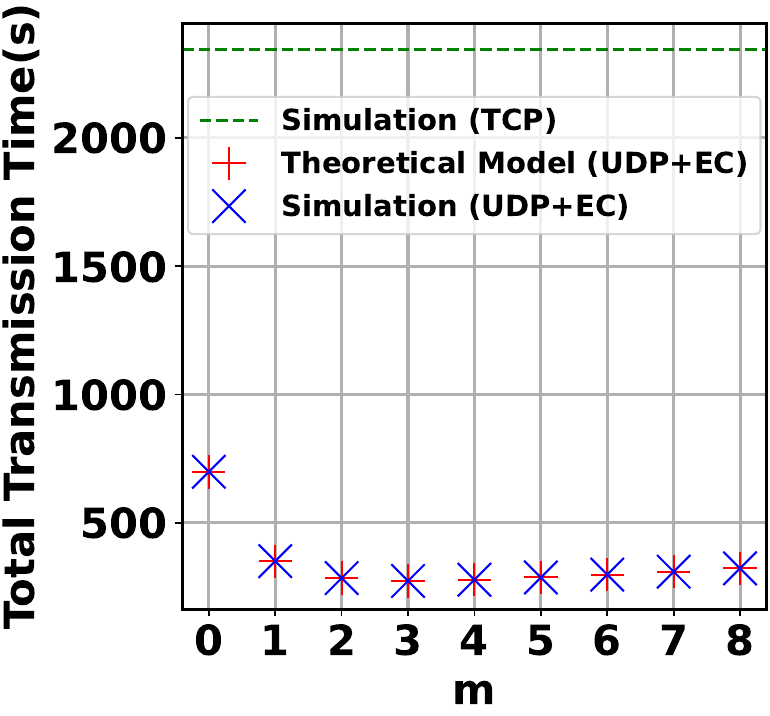}}
  \subfigure[Time-varying packet loss rate]{ 
  \label{fig:lambda_varying_time}
    \includegraphics[width=.23\linewidth]{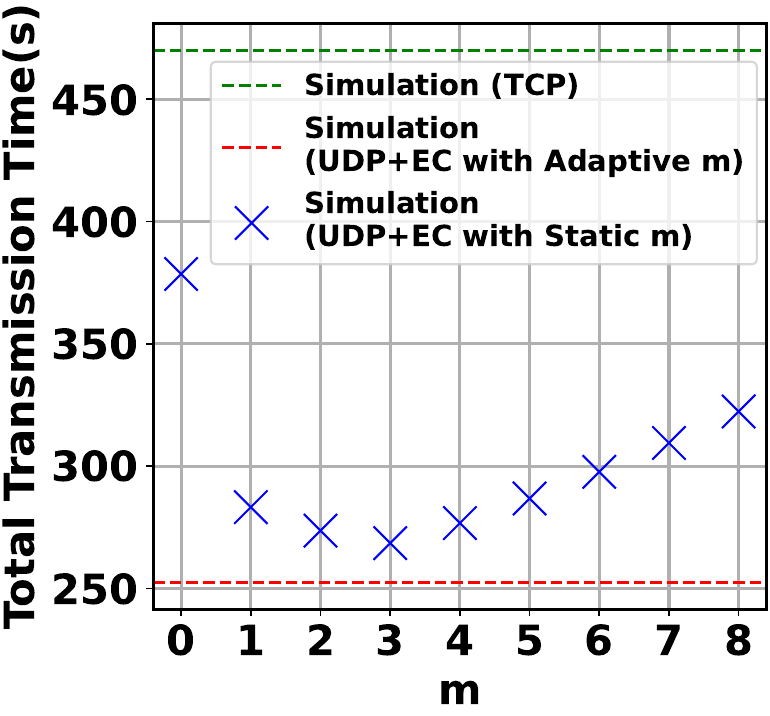}}
  \caption{Total time for transferring data with guaranteed error bound under different packet loss rates}
  \label{fig:with_guaranteed_error}
\end{figure*}

In this evaluation, we focus on the data transfer process where an error-bound must be guaranteed for the received data. Assuming the user specifies that the error in the received data must remain below $0.00001$, all four levels of the refactored Nyx cosmology simulation data must be successfully received to meet this requirement.

While packet loss rates in real-world wide-area networks vary dynamically over time, we first conducted simulation-based evaluations under static packet loss conditions to validate the analytical models developed in~\autoref{sec:model}. 
In these baseline experiments, we simulated the data transfer of four hierarchical levels using both the TCP protocol and a UDP-based fault-tolerant protocol. The UDP-based approach enhances traditional UDP transmission with erasure coding and passive retransmission. In this configuration, each fault-tolerance group (FTG) includes a fixed number of parity fragments, denoted as $m$, which are used to recover lost packets without requiring end-to-end acknowledgment. For each possible value of $m$, we performed multiple simulations and recorded the total time required to complete the transfer across all levels. In parallel, we computed the expected total transfer time, $E[T_{total}]$, using the analytical expressions derived in~\autoref{sec:model}, allowing for direct comparison between simulated and theoretical results.

The evaluation results are presented in~\autoref{fig:lambda_19_time}--~\autoref{fig:lambda_957_time}, providing the following insights: 1) When the TCP protocol is used, the data transmission time increases significantly as the packet loss rate rises. 2) For the UDP-based protocol that integrates erasure coding and passive retransmission, the total transmission time derived from the theoretical equations in~\autoref{sec:model} aligns closely with the simulation results. This alignment confirms that our model effectively captures the impact of packet loss, parity fragment generation, and retransmission on overall transmission performance. 3) When the packet loss rate is low (\autoref{fig:lambda_19_time}), increasing the number of parity fragments per FTG negatively affects performance due to the added overhead. However, as the packet loss rate increases (\autoref{fig:lambda_383_time} and~\autoref{fig:lambda_957_time}), there exists an optimal value of $m$ that minimizes the total transmission time.

When the packet loss rate changes over time, a static fault-tolerance configuration often fails to achieve optimal transmission performance. To demonstrate this, we conducted simulation-based evaluations to assess the effectiveness of the proposed data transfer protocols in dynamically adjusting fault-tolerance configurations in response to time-varying packet loss rates. We compare the proposed protocols with two other baseline protocols: the TCP protocol, a UDP-based protocol incorporating erasure coding with a static number of parity fragments per FTG. As illustrated in~\autoref{fig:lambda_varying_time}, the adaptive protocol consistently outperforms the other two approaches under time-varying packet loss rates. Notably, the adaptive protocol completes the data transmission in 252.4 seconds, reducing the transmission time by about 20 seconds compared to the minimum time achieved by the static fault-tolerance configuration.

\subsubsection{Data Transfer with Guaranteed Time}
\begin{figure*}[htbp]
  \centering 
  \subfigure[Low packet loss rate]{ 
  \label{fig:lambda_19_error}
    \includegraphics[width=.48\textwidth]{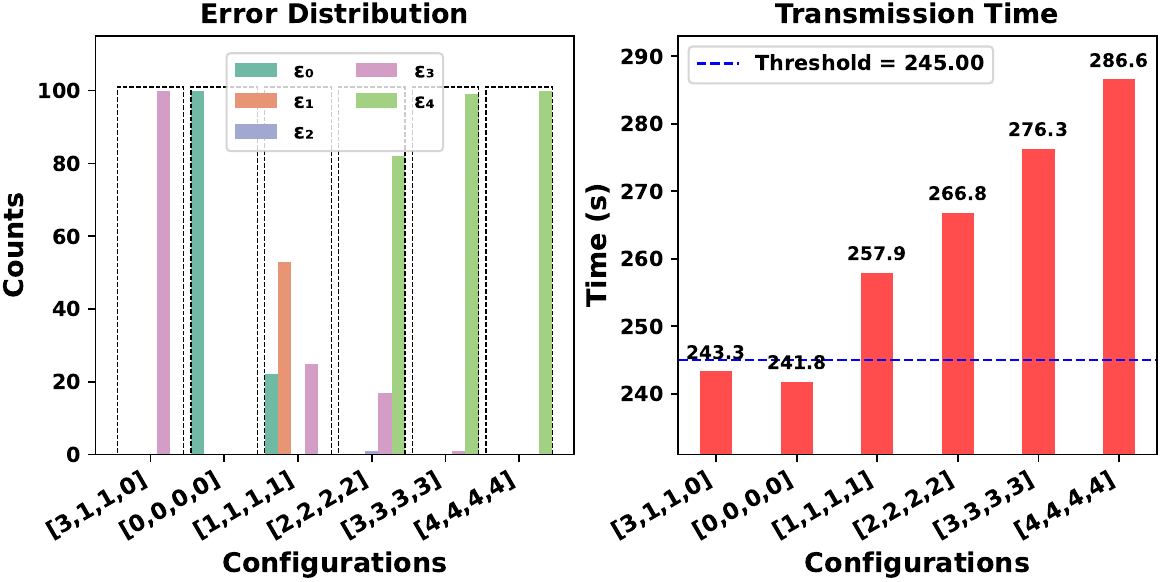}}
  \subfigure[Medium packet loss rate]{ 
  \label{fig:lambda_383_error}
    \includegraphics[width=.48\textwidth]{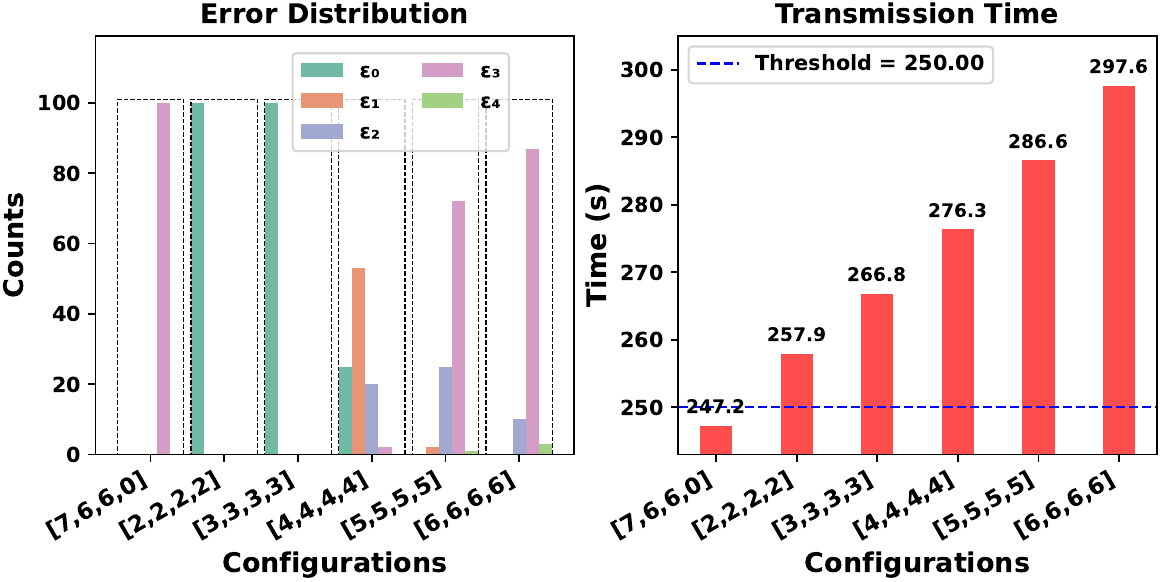}}
  \subfigure[High packet loss rate]{ 
  \label{fig:lambda_957_error}
    \includegraphics[width=.48\textwidth]{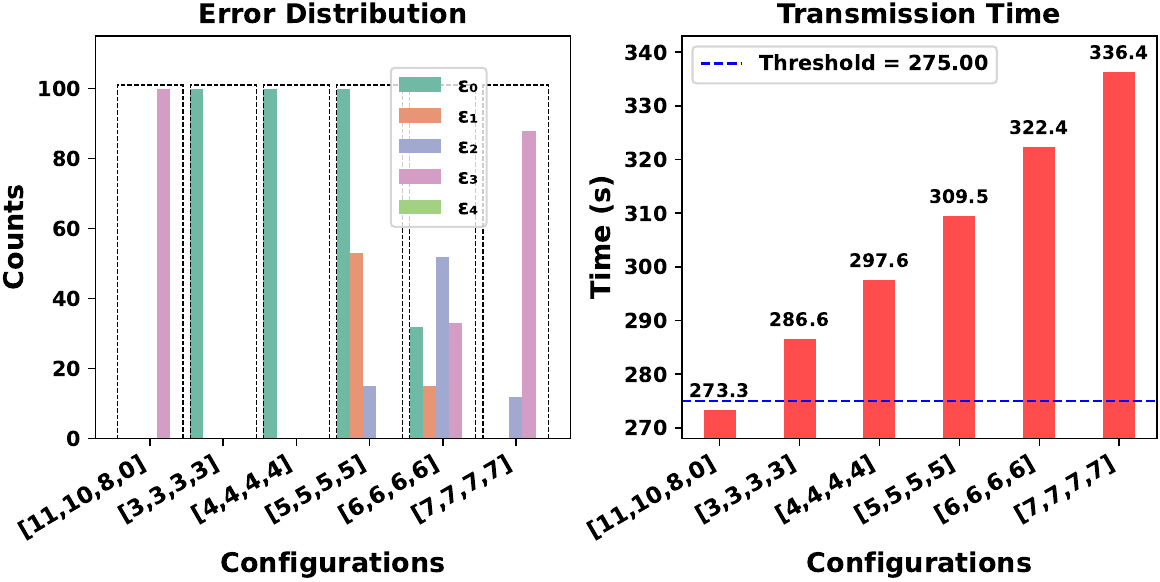}}
  \subfigure[Time-varying packet loss rate]{ 
  \label{fig:lambda_varying_error}
    \includegraphics[width=.48\textwidth]{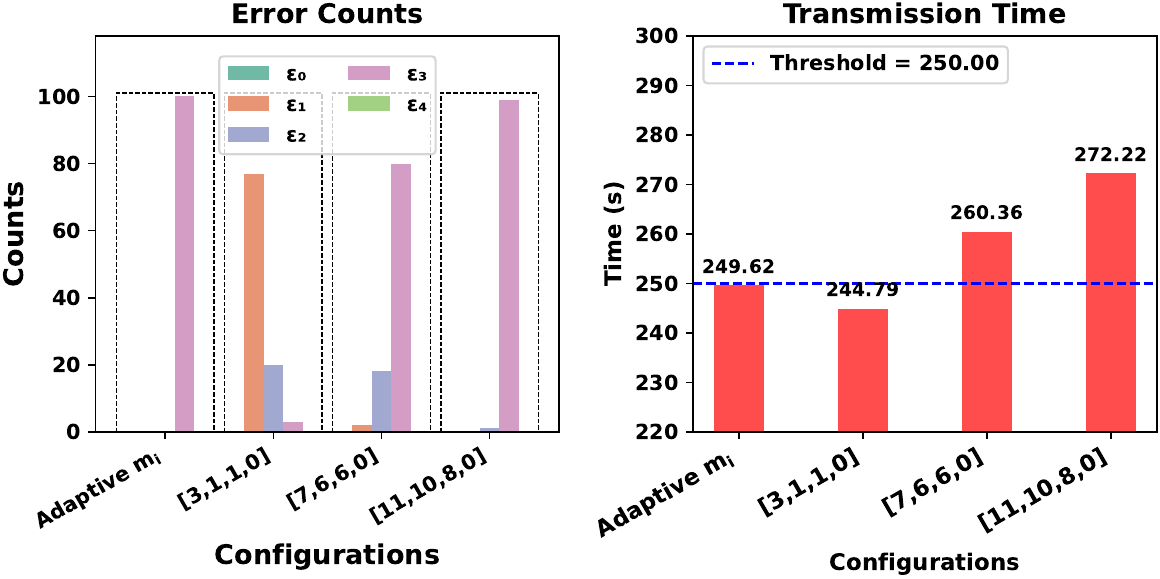}}
  \caption{Error bounds of data received within guaranteed transmission time under different packet loss rates}
  \label{fig:lambda_guaranteed_time}
\end{figure*}

In this evaluation, we focus on the data transfer process under a guaranteed transmission time constraint, which allows users to receive data within a specified time frame while accepting that the error bound of the received data may not be strictly guaranteed. This scenario reflects practical use cases where timely data delivery is prioritized over perfect accuracy.

We first conducted simulations under static packet loss rates to analyze the system’s behavior and validate the optimization model introduced in~\autoref{epsilon_opt}. As shown in\autoref{fig:lambda_19_time}--~\autoref{fig:lambda_957_time}, the minimum transmission times required to complete the transfer of all refactored data levels were 247.2 seconds under low packet loss ($\lambda = 20$), 260.6 seconds under medium loss ($\lambda = 400$), and 276.3 seconds under high loss ($\lambda = 1000$). Since users in this mode are willing to trade data accuracy for faster completion, we defined transmission time constraints $\tau$ of 245 seconds, 250 seconds, and 275 seconds for the low, medium, and high loss scenarios, respectively. The optimization results indicate that minimizing the reconstruction error within these time limits requires adjusting the number of parity fragments per FTG across levels. Specifically, the optimal configurations were $m_1=3$, $m_2=1$, $m_3=1$, and $m_4=0$ for $\lambda = 20$; $m_1=7$, $m_2=6$, $m_3=6$, and $m_4=0$ for $\lambda = 400$; and $m_1=11$, $m_2=10$, $m_3=8$, and $m_4=0$ for $\lambda = 1000$.

To evaluate the effectiveness of these configurations, we performed 100 independent simulation runs per scenario to account for the stochastic nature of packet loss and data corruption. In each run, we recorded the final error bounds of the received data, such as $\varepsilon_0$, $\varepsilon_1$, and $\varepsilon_3$. Aggregating results from 100 runs allows us to quantify the likelihood of achieving different levels of reconstruction accuracy under the specified time constraints.
To assess the benefit of the optimized settings, we compared them with uniform configurations, where the number of parity fragments per FTG was fixed across all four levels. As shown in~\autoref{fig:lambda_19_error}--~\autoref{fig:lambda_957_error}, the optimized configurations consistently met the transmission time constraint $\tau$ while maintaining a high probability of low error bounds (e.g., $\varepsilon_3$). In contrast, uniform configurations either satisfied the time constraint but resulted in high error bounds such as $\varepsilon_0$, or exceeded the transmission time limit. Note that in each run, simulations continued even if the transmission time exceeded the threshold. With a strict cut-off, transfers using uniform configurations would have a much higher likelihood of receiving data with higher error bounds than shown in~\autoref{fig:lambda_19_error}--~\autoref{fig:lambda_957_error}.
Next, we conducted simulations to evaluate the proposed adaptive protocol in~\autoref{alg:transfer_with_time}, which transfers data within a guaranteed time constraint under time-varying packet loss rates. Based on prior experiments, the minimum transmission time required to transfer all data levels using the protocol in~\autoref{alg:transfer_with_err_bound} under the same conditions was 252.4 seconds. Accordingly, we set the transmission time constraint to 250 seconds to reflect the scenario in which the user is willing to trade a small degree of accuracy in the received data for faster delivery. Because the packet loss rate changes over time, no static fault-tolerance configuration can simultaneously minimize reconstruction error and satisfy the time constraint across all conditions. To provide a baseline for comparison, we also conducted simulations using the optimal static configurations previously derived for constant-loss scenarios (\autoref{fig:lambda_19_error}–\autoref{fig:lambda_957_error}) under the same time-varying loss patterns.

We performed 100 simulation runs using these static fault-tolerance configurations and another 100 runs using the adaptive protocol described in~\autoref{alg:transfer_with_time}. For each run, we recorded the number of instances in which the received data satisfied different error bounds. As shown in~\autoref{fig:lambda_varying_error}, the adaptive protocol consistently maintained transmission completion within the specified time constraint while achieving a high probability of low reconstruction error. In contrast, the static configurations either met the time constraint but yielded data with higher error bounds or failed to complete within the required time.

\subsection{Real Network Testing}

\subsubsection{C++ Implementation of Adaptive Transfer Protocols} 
To evaluate the performance of data transmission over real wide-area networks, we developed a prototype of the proposed adaptive data transfer protocols, as described in~\autoref{alg:transfer_with_err_bound} and~\autoref{alg:transfer_with_time}, using C++. Our implementation leverages the Boost.Asio~\cite{Boost.Asio} library for efficient UDP-based data transmission. Additionally, we utilize the Protocol Buffers (Protobuf) library~\cite{Protobuf} to structure packet contents, embedding essential erasure coding metadata—such as the level and FTG each fragment belongs to, as well as its corresponding redundancy level ($m$)—alongside the actual data or parity fragment. The prototype, comprising over 1,600 lines of C++ code, provides a scalable and flexible foundation for evaluating the proposed adaptive transfer mechanisms in real-network environments.

\subsubsection{Experimental Setup}
\begin{figure}[h] \centering
\includegraphics[width=0.9\columnwidth]{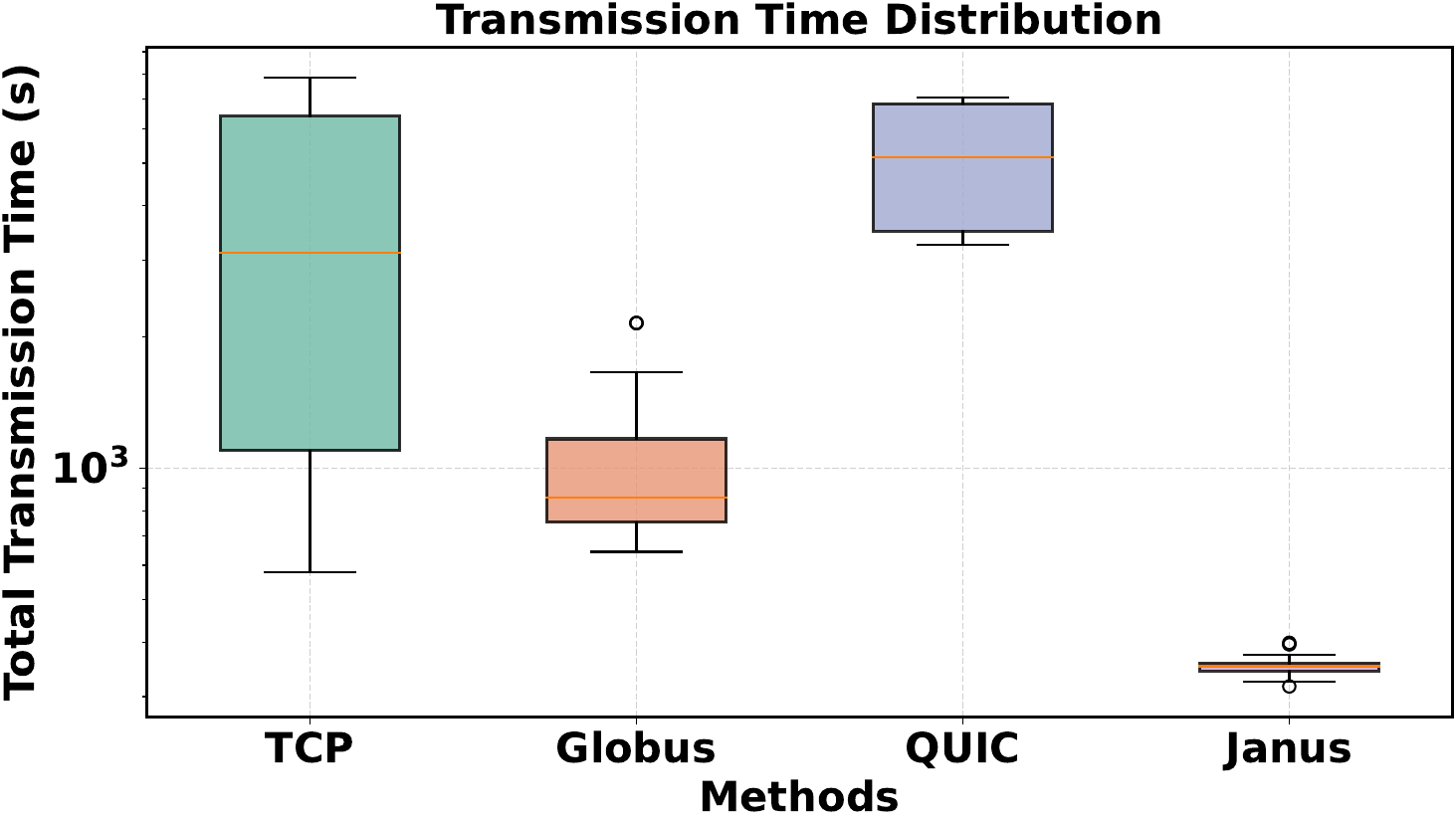}
\caption{Total time for transferring data with guaranteed error bound using different data transfer protocols}
\label{fig:real_test_guaranteed_error}
\end{figure}

\begin{figure}[h] \centering
\includegraphics[width=0.9\columnwidth]{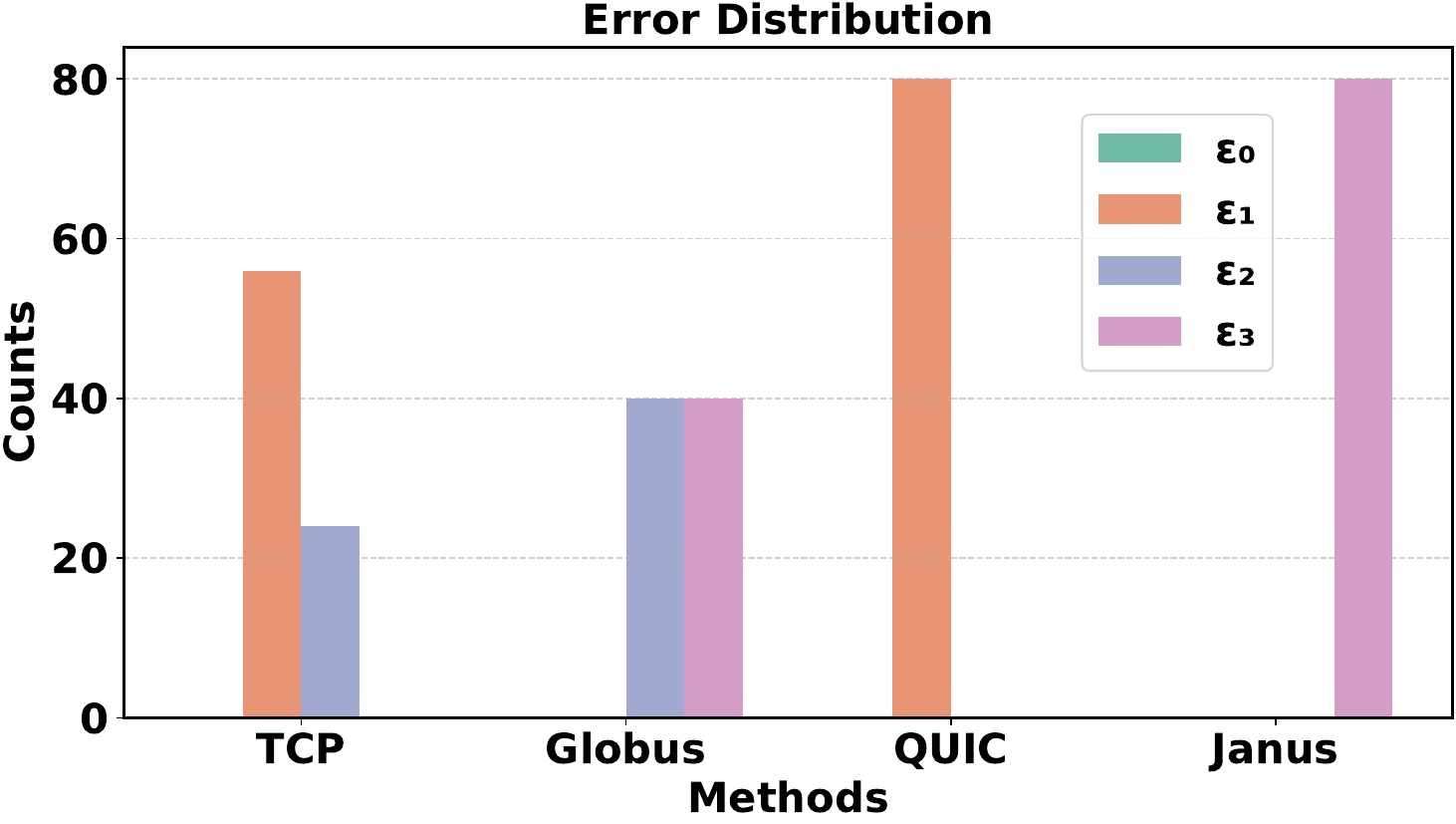}
\caption{Error bounds of data received within guaranteed transmission time}
\label{fig:real_test_guaranteed_time}
\end{figure}
In each test run, we performed a series of measurements. First, we used the native TCP protocol to transfer all levels of the refactored Nyx dataset from the sender to the receiver and recorded the total transmission time. Next, we employed the Globus service~\cite{Globus} to transfer the same data and measured its transmission time. We then adopted the QUIC protocol~\cite{10.1145/3098822.3098842}, the state-of-the-art UDP-based transport protocol with built-in fault-tolerance mechanisms for handling packet loss. Afterward, we applied the adaptive protocol described in~\autoref{alg:transfer_with_err_bound} to transmit the data while ensuring an error bound guarantee of $\varepsilon_4$ and requiring successful receipt of all data levels. Finally, we used the adaptive protocol in~\autoref{alg:transfer_with_time} to transfer the data within a fixed time constraint set to 90\% of the transmission time achieved by the error-bound-guaranteed protocol. To account for variability in packet loss rates, all experiments were repeated 80 times at different times of day and on different days.

\subsubsection{Testing Results}
As shown in~\autoref{fig:real_test_guaranteed_error}, when all levels of the Nyx data are required to be received, the transmission times of the native TCP, Globus, and QUIC protocols vary considerably across test runs, highlighting their sensitivity to packet loss. In contrast, our adaptive data transfer protocol not only achieves substantially lower transmission times but also exhibits much greater stability across runs. Moreover, as presented in~\autoref{fig:real_test_guaranteed_time}, when employing the adaptive protocol described in~\autoref{alg:transfer_with_time} to satisfy the specified time constraints, all 100 test runs successfully delivered data within the error bound $\varepsilon_3$. In comparison, when the other protocols are used, the received data is more likely to exceed this error bound when the same time limit is reached. 

\section{Related Work}

UDP, a connectionless and low-latency transport protocol, lacks built-in error correction mechanisms~\cite{postel1980rfc}, making it particularly susceptible to packet loss in high-performance and wide-area network environments. To address this limitation, many studies have explored lightweight and adaptive flow control mechanisms to improve the reliability of UDP-based data transfer.
One notable approach is Performance-Adaptive UDP (PA-UDP), a high-speed data transfer protocol proposed by~\cite{eckart2008performance}. PA-UDP enables dynamic flow control based on the sending system’s performance characteristics, allowing UDP packets to be sent asynchronously without blocking TCP transmission. 
The SDUDP framework~\cite{wang2017sdudp} takes advantage of Software-Defined Networking (SDN)'s capabilities, such as centralized flow control and real-time packet monitoring, to achieve reliable UDP data transfer. 
Beyond SDUDP, various studies have explored Reliable UDP (RUDP) protocols designed to enhance UDP’s performance while maintaining its low-latency advantage over TCP. 
These protocols typically introduce mechanisms such as selective acknowledgments, forward error correction (FEC), and retransmission strategies to mitigate packet loss. However, many RUDP implementations require modifications to end-host systems, making deployment challenging in large-scale networks~\cite{tran2007reliable, masirap2016evaluation, wang2006reliable}. In addition, QUIC (Quick UDP Internet Connections), a transport layer protocol developed by Google, uses mechanisms like congestion control and loss recovery to enhance packet delivery~\cite{10.1145/3098822.3098842}.

To further optimize UDP-based data transfer, research has focused on detecting and predicting packet loss more efficiently. For instance, Wu et al.~\cite{wu2022lossdetection} proposed a novel framework, LossDetection, for real-time packet loss detection in both TCP and UDP traffic. LossDetection employs Feature-Sketch, a technique that extracts key features from packet streams for efficient and scalable loss identification. By detecting packet loss patterns dynamically, this method enables more effective retransmission strategies.
Another approach to mitigating packet loss involves predictive modeling. Giannakou et al.~\cite{giannakou2020machine} developed a lightweight Random Forest Regression (RFR) model to predict retransmissions in scientific data transfers. Their model builds upon existing loss detection techniques by forecasting packet loss trends and proactively adjusting transmission parameters to minimize unnecessary retransmissions. 

Beyond optimizing flow control mechanisms, another type of approach to improving UDP reliability involves introducing redundancy into transmitted data to tolerate packet losses. A fundamental technique used in these approaches is erasure coding, which 
has been extensively studied and implemented to enhance fault tolerance in both data storage and transmission. Foundational work by Reed and Solomon introduced Reed-Solomon codes~\cite{wicker1999reed}, which remain a gold standard for error correction. 
Several studies have explored the integration of erasure coding with UDP-based data transfer to improve reliability. For instance, Byers et al.~\cite{byers1998digital} demonstrated that incorporating Reed-Solomon codes into UDP streaming significantly enhances packet recovery, mitigating losses common in lossy network conditions. Further research has investigated the use of fountain codes, such as Raptor codes, over UDP to facilitate reliable high-bandwidth data transfers. Mtibaa and Bestavros proposed RC-UDP~\cite{mtibaa2018rc}, a framework that leverages Raptor coding over UDP to achieve efficient and reliable data transmission. Their findings indicate that this approach effectively addresses packet loss without incurring significant overhead. 

Despite these advancements, several challenges remain unaddressed. One major challenge is the computational and bandwidth overhead introduced by fault-tolerance techniques such as erasure coding. This overhead can be particularly significant when transferring large-scale scientific datasets over wide-area networks. To address this, we propose integrating erasure coding with lossy compression, effectively reducing the data volume while maintaining a controllable level of fidelity. Another key challenge is the dynamic nature of network conditions, particularly fluctuating packet loss rates, which can lead to suboptimal performance when using static fault-tolerance configurations. Existing static approaches fail to adapt to real-time variations in network reliability, often resulting in either excessive redundancy or insufficient protection against data loss. To overcome this, our work introduces an adaptive optimization framework that dynamically adjusts erasure coding parameters based on real-time network conditions. 
\section{Conclusion}
\label{sec:conclusion}

In this paper, we introduced \sys, a resilient and adaptive data transmission approach designed to enhance the efficiency and reliability of scientific data transfers over wide-area networks. By leveraging UDP, integrating erasure coding for fault tolerance, and incorporating error-bounded lossy compression to reduce overhead, \sys enables users to optimize data transmission time while preserving accuracy. Furthermore, its ability to dynamically adjust erasure coding parameters based on real-time network conditions ensures sustained performance under fluctuating environments. Our extensive evaluations, including both simulations and real-network experiments, demonstrate that \sys significantly improves transfer efficiency while maintaining data integrity. As cross-facility scientific workflows continue to expand, solutions like \sys will play a crucial role in enabling timely and scalable data sharing, ultimately accelerating scientific discovery.

\section*{Acknowledgment}
Research was sponsored by the Army Research Laboratory and was accomplished under Cooperative Agreement Number W911NF-23-2-0224. The views and conclusions contained in this document are those of the authors and should not be interpreted as representing the official policies, either expressed or implied, of the Army Research Laboratory or the U.S. Government. The U.S. Government is authorized to reproduce and distribute reprints for Government purposes notwithstanding any copyright notation herein.

\bibliographystyle{IEEEtran}
\bibliography{references}

\end{document}